\documentclass[preprint,aps,nofootinbib]{revtex4}
\usepackage{graphicx}
\usepackage{epstopdf}
\usepackage{subfigure}
\usepackage{textcomp}
\def\bkR{{\rm I\kern-.17em R}}
\def\bkC{{\rm \kern.24em \vrule width.05em height1.4ex depth-.05ex \kern-.26em C}}
\def\to{\rightarrow}

\def\be{\beta}

\def\frac#1#2{{\textstyle{{#1}\over {#2}}}}

\def\lsim{\mathrel{\rlap{\lower4pt\hbox{\hskip1pt$\sim$}}
    \raise1pt\hbox{$<$}}}
\def\gsim{\mathrel{\rlap{\lower4pt\hbox{\hskip1pt$\sim$}}
    \raise1pt\hbox{$>$}}}
\def\sqr#1#2{{\vcenter{\vbox{\hrule height.#2pt
         \hbox{\vrule width.#2pt height#1pt \kern#1pt
         \vrule width.#2pt}
         \hrule height.#2pt}}}}

\def\laq{\raise 0.4 ex \hbox{$<$}\kern -0.8 em\lower 0.62 ex\hbox{$\sim$}}
\def\gaq{\raise 0.4 ex \hbox{$>$}\kern -0.7 em\lower 0.62 ex\hbox{$\sim$}}

\def\be{\begin{equation}}
\def\ee{\end{equation}}
\def\ba{\begin{eqnarray}}
\def\ea{\end{eqnarray}}

\def\dalemb#1#2{{\vbox{\hrule height.#2pt
        \hbox{\vrule width.#2pt height#1pt \kern#1pt \vrule width.#2pt}
        \hrule height.#2pt}}}

\def\dalemb#1#2{{\vbox{\hrule height.#2pt
        \hbox{\vrule width.#2pt height#1pt \kern#1pt \vrule width.#2pt}
        \hrule height.#2pt}}}

\def\gtorder{\mathrel{\raise.3ex\hbox{$>$}\mkern-14mu
             \lower0.6ex\hbox{$\sim$}}}
\def\ltorder{\mathrel{\raise.3ex\hbox{$<$}\mkern-14mu
             \lower0.6ex\hbox{$\sim$}}}

\begin{document}

\rightline{DF/IST-7.2009}
\rightline{December 2009}

\title{Black Holes and Phase-Space Noncommutativity}

\author{Catarina Bastos\footnote{Also at Instituto de Plasmas e Fus\~ao Nuclear,
IST. E-mail: catarina.bastos@ist.utl.pt},
Orfeu Bertolami\footnote{Also at Instituto de Plasmas e Fus\~ao Nuclear,
IST. E-mail: orfeu@cosmos.ist.utl.pt}}

\vskip 0.3cm

\affiliation{Departamento de F\'\i sica, Instituto Superior T\'ecnico \\
Avenida Rovisco Pais 1, 1049-001 Lisboa, Portugal}

\author{Nuno Costa Dias\footnote{Also at Grupo de F\'{\i}sica Matem\'atica, UL,
Avenida Prof. Gama Pinto 2, 1649-003, Lisboa, Portugal. E-mail: ncdias@meo.pt}, Jo\~ao Nuno Prata\footnote{Also at Grupo de F\'{\i}sica Matem\'atica, UL,
Avenida Prof. Gama Pinto 2, 1649-003, Lisboa, Portugal. E-mail: joao.prata@mail.telepac.pt}}

\vskip 0.3cm

\affiliation{Departamento de Matem\'{a}tica, Universidade Lus\'ofona de
Humanidades e Tecnologias \\
Avenida Campo Grande, 376, 1749-024 Lisboa, Portugal}


\vskip 1cm

\begin{abstract}

\vskip 1cm

{We use the solutions of the noncommutative Wheeler-De Witt equation arising from a Kantowski-Sachs cosmological model to compute thermodynamic properties of the Schwarzschild black hole. We show that the noncommutativity in the momentum sector introduces a quadratic term in the potential function of the black hole minisuperspace model. This potential has a local minimum and thus the partition function can be computed by resorting to a saddle point evaluation in the neighbourhood of the minimum. The thermodynamics of the black hole is derived and the corrections to the usual Hawking temperature and entropy exhibit a dependence on the momentum noncommutative parameter, $\eta$. Moreover, we study the $t=r=0$ singularity in the noncommutative regime and show that in this case the wave function of the system vanishes in the neighbourhood of $t=r=0$.}

\end{abstract}

\maketitle

\section{Introduction}

Black Holes (BHs) are one of the most puzzling legacies of general relativity. Their existence implies that space-time is geodesically incomplete and has at least one singular point where physical quantities are divergent. Furthermore, semiclassical considerations reveal, rather surprisingly, that BHs radiate and have an associated thermodynamics. It is therefore natural to expect that a quantum gravity theory might allow for an underlying statistical mechanical explanation of the thermodynamical properties of BHs and shed some light on the prevalence and possible evolution of space-time singularities.

In this respect, given that a quantum gravity theory is not within grasp, a quantum cosmology approach based on the mininsuperspace approximation might bring some insight about the prevalence of singularities under quantum conditions. Furthermore, the evidence that noncommutative features are present in string theory, one of the most discussed quantum gravity proposals, suggests that some noncommutativity structure should be equally included into the quantum cosmology approach. This means that one should seek for a noncommutative version of the Wheeler-De Witt equation. This generalization has been carried out in the context of the Kantowski-Sachs (KS) cosmological model for the most general (canonical) noncommutative algebra, namely the one that includes noncommutativity of the momentum variables as well \cite{Bastos2}. An additional bonus of this approach is that the KS model can be used to study the interior region of a Schwarzschild BH.

Thus, the aim of the present work is to use the phase-space noncommutative generalization of the KS cosmological model developed in Ref. \cite{Bastos2} to examine the interior of a Schwarzschild BH. As we shall see this will allow for a proper treatment of the BH thermodynamics, contrary to what happens when momentum noncommutativity is absent, and for an analysis of the impact of the quantum and noncommutative effects on the issue of singularities.

Let us start our discussion with the description of a Schwarzschild BH. General relativity allows for solutions where the causal structure of space-time changes at different regions of space-time. A well known example is the Schwarzschild BH, described by the following metric,
\be\label{eq0.1}
ds^2=-\left(1-{2M\over r}\right)dt^2+\left(1-{2M\over r}\right)^{-1}dr^2+r^2(d\theta^2+\sin^2\theta d\varphi^2)~,
\ee
where $r$ is the radial coordinate. The fact that the first component of the metric vanishes at $r=2M$ defines the horizon of events for the Schwarzschild BH. For $r<2M$ the time and radial coordinates are interchanged ($r\leftrightarrow t$) so that space-time is described by the metric
\be\label{eq0.2}
ds^2=-\left({2M\over t}-1\right)^{-1}dt^2+\left({2M\over t}-1\right)dr^2+t^2(d\theta^2+\sin^2\theta d\varphi^2)~.
\ee
This means that an isotropic metric turns into an anisotropic one, implying that the interior of a Schwarzschild BH can be described by an anisotropic cosmological space-time. Indeed, the metric (\ref{eq0.2}) can be mapped to the metric of the KS cosmological model \cite{Anderson,Kantowski}, which, in the Misner parametrization, can be written as
\be\label{eq1.4}
ds^2=-N^2dt^2+e^{2\sqrt{3}\beta}dr^2+e^{-2\sqrt{3}\beta}e^{-2\sqrt{3}\Omega}(d\theta^2+\sin^2{\theta}d\varphi^2)~,
\ee
where $\Omega$ and $\beta$ are scale factors, and $N$ is the lapse function. The following identification for $t<2M$,
\be\label{eq0.3}
N^2=\left({2M\over t}-1\right)^{-1} \quad, \quad
e^{2\sqrt{3}\beta} = \left({2M\over t}-1\right) \quad , \quad
e^{-2\sqrt{3}\beta}e^{-2\sqrt{3}\Omega}=t^2~,
\ee
allows for turning the metric Eq. (\ref{eq1.4}) into the metric Eq. (\ref{eq0.2}) away from the horizon $t=r=2M$, the coordinate singularity.

In what follows we shall assume that, at the quantum level, the interior of the BH is described by the phase-space noncommutative extension of the quantum KS cosmological model. In this setting, the noncommutative Wheeler-De Witt (NCWDW) equation has been recently obtained and its solutions calculated explicitly \cite{Bastos2}. As we shall see, the introduction of momentum noncommutativity changes the WDW equation and has a significant impact on the functional form of its solutions. The new WDW equation allows for consistently calculating the partition function for the Schwarzschild BH through the Feynman-Hibbs procedure \cite{Feynman} and then on all the relevant thermodynamic quantities. This, as we will see, is in contrast with what happens in the commutative and configuration space noncommutative cases. Moreover, the solutions of the NCWDW equation provide a new insight into the problem of the BH singularity. We will show that these solutions vanish in the neighbourhood of $t=r=0$, but that this does not necessarily imply a vanishing probability of finding the system at the singularity. We conclude that the canonical phase-space noncommutativity discussed in this work is insufficient to ensure the avoidance of singularities, but quite interestingly, our results strongly suggest that the singularity may be removed through other, slightly different, forms of momentum noncommutativity.

This paper is organized as follows: We start by recalling the main features of phase-space noncommutative quantum mechanics (NCQM) and quantum cosmology in section II. In section III, we use the solutions of the phase-space NCWDW equation to obtain the relevant thermodynamical properties of the Schwarzschild BH through the Feynman-Hibbs procedure. In section IV we discuss the implications of our results for the BH singularity issue. Section V contains our conclusions.

\section{Noncommutative Phase-Space Quantum Cosmology}

It is believed that the theory of quantum gravity will determine the ultimate structure of space-time at the Planck scale. It has been pointed out that at this scale space-time might have a noncommutative structure, as suggested, for instance by string theory \cite{Connes, Seiberg}. In this work we use units such that $\hbar = c = k =G=1$, which imply for the Planck length and mass $l_P =M_P=1$.

At the quantum mechanical level, noncommutativity has been extensively discussed \cite{Bertolami1,Bertolami2,Zhang, Acatrinei,Bastos1,Bastos4}. In this instance, one considers canonical extensions of the Heisenberg-Weyl algebra, where time is assumed to be a commutative parameter and the theory is set in a $2d$-dimensional phase-space of operators with noncommuting position and momentum variables. The noncommutative algebra reads,
\be\label{eq1.1}
\left[\hat q_i, \hat q_j \right] = i\theta_{ij} \hspace{0.2 cm}, \hspace{0.2 cm} \left[\hat q_i, \hat p_j \right] = i  \delta_{ij} \hspace{0.2 cm},
\hspace{0.2 cm} \left[\hat p_i, \hat p_j \right] = i \eta_{ij} \hspace{0.2 cm},  \hspace{0.2 cm} i,j= 1, ... ,d
\ee
where $\eta_{ij}$ and $\theta_{ij}$ are antisymmetric real constant ($d \times d$) matrices and $\delta_{ij}$ is the identity matrix. The extended algebra is related to the standard Heisenberg-Weyl algebra:
\be\label{eq1.2}
\left[\hat R_i, \hat R_j \right] = 0 \hspace{0.2 cm}, \hspace{0.2 cm} \left[\hat R_i, \hat \Pi_j \right]
= i \hbar \delta_{ij} \hspace{0.2 cm}, \hspace{0.2 cm} \left[\hat \Pi_i, \hat \Pi_j \right] = 0 \hspace{0.2 cm},
\hspace{0.2 cm} i,j= 1, ... ,d ~,
\ee
by a class of linear (non-canonical) transformations:
\be\label{eq1.3}
\hat q_i = \hat q_i \left(\hat R_j , \hat \Pi_j \right) \hspace{0.2 cm},\hspace{0.2 cm}
\hat p_i = \hat p_i \left(\hat R_j , \hat \Pi_j \right)~.
\ee
In the mathematics literature this mapping is usually referred to as Darboux (D) map \cite{Bastos1}.

A phase-space noncommutative extension of a quantum cosmology was studied in Ref. \cite{Bastos2}. The presence of at least two scale factors is necessary to consider such an extension and the KS cosmological model, based on the metric Eq. (\ref{eq1.4}), is a most natural candidate. In this case, we impose the following (noncommutative) classical Poisson algebra on the scale factors $\beta$, $\Omega$ and their conjugate momenta $P_{\beta}$, $P_{\Omega}$
  \be\label{eq1.4.1}
\begin{array}{l l}
\left\{\Omega, P_{\Omega} \right\} =1 & \left\{\beta, P_{\beta} \right\} =1\\
& \\
\left\{\Omega, \beta \right\} = \theta & \left\{P_{\Omega}, P_{\beta} \right\} =\eta~.
\end{array}
\ee
The remaining brackets vanish.

Following the ADM procedure and taking $\Omega, \beta$ as configuration variables, one can derive the Hamiltonian constraint for this system,
\be\label{eq1.5}
H=N{\cal H}=Ne^{\sqrt{3}\beta+2\sqrt{3}\Omega}\left[-{P_{\Omega}^2\over24}+{P_{\beta}^2\over24}-2e^{-2\sqrt{3}\Omega}\right]~.
\ee
The lapse function for the Hamiltonian problem associated to metric Eq. (\ref{eq1.4}), $N$, is chosen as $N=24e^{-\sqrt{3}\beta-2\sqrt{3}\Omega}$ \cite{Bastos2}. With this choice, the classical equations of motion simplify drastically. For the algebra (\ref{eq1.4.1}), these read:
\ba\label{eq1.6}
&&\dot{\Omega}=-2P_{\Omega}~,\hspace{0.5cm}(a)\nonumber\\
&&\dot{P_{\Omega}}=2\eta P_{\beta}-96\sqrt{3}e^{-2\sqrt{3}\Omega}~,\hspace{0.5cm}(b)\nonumber\\
&&\dot{\beta}=2P_{\beta}-96\sqrt{3}\theta e^{-2\sqrt{3}\Omega}~,\hspace{0.5cm}(c)\nonumber\\
&&\dot{P_{\beta}}=2\eta P_{\Omega}~.\hspace{0.5cm}(d)
\ea
An analytical solution of this system is beyond reach, given the entanglement among the four variables. However, a numerical solution can be obtained and used to provide predictions for several relevant physical quantities \cite{Bastos2}. Despite the difficulty to obtain analytic solutions one notices that Eqs. (\ref{eq1.6}a) and (\ref{eq1.6}d) yield a constant of motion
\be\label{eq1.7}
\dot{P_{\beta}}=-\eta(-2P_{\Omega})=-\eta\dot{\Omega}\Rightarrow P_{\beta}+\eta\Omega=C~,
\ee
which will play an important role in solving the phase-space NCWDW equation.

The canonical quantization of the classical Hamiltonian constraint, ${\cal H}\approx 0$, based on the ordinary Heisenberg-Weyl algebra, yields the commutative WDW equation for the wave function of the universe. For the simplest factor ordering of operators this equation reads
\be\label{eq1.8}
\left[- \hat P^2_{\Omega}+ \hat P^2_{\beta}-48e^{-2\sqrt{3} \hat{\Omega}}\right]\psi(\Omega,\beta)=0~.
\ee
where $\hat P_{\Omega}=-i \frac{\partial }{\partial \Omega}$, $\hat P_{\beta}=-i \frac{\partial }{\partial \beta}$ are the fundamental momentum operators conjugate to $\hat{\Omega} = \Omega$ and $\hat{\beta} = \beta$, respectively. Here, on the other hand, we shall perform the quantization of the classical algebra (\ref{eq1.4.1}). We thus obtain the following noncommutative Heisenberg-Weyl algebra:
 \be\label{eq1.8.1}
\begin{array}{l l}
\left[\hat{\Omega}, \hat{P}_{\Omega} \right] =i & \left[\hat{\beta}, \hat{P}_{\beta} \right] =i\\
& \\
\left[\hat{\Omega}, \hat{\beta} \right] = i \theta & \left[ \hat{P}_{\Omega}, \hat{P}_{\beta} \right] = i \eta~.
\end{array}
\ee
The remaining commutators vanish.

The simplest way to implement algebra (\ref{eq1.8.1}) in this context is through a suitable D map (i.e. a non-unitary linear transformation mapping the noncommutative algebra into the standard Heisenberg-Weyl algebra) \cite{Bastos2}
\ba\label{eq1.9}
\hat{\Omega} =\lambda \hat{\Omega}_{c}-{\theta\over2\lambda} \hat P_{\beta_c} \hspace{0.2cm} , \hspace{0.2cm} \hat{\beta} = \lambda \hat{\beta}_{c} + {\theta\over2\lambda} \hat P_{\Omega_c}~,\nonumber\\
\hat P_{\Omega}= \mu \hat P_{\Omega_c} + {\eta\over2\mu} \hat{\beta}_{c} \hspace{0.2cm} , \hspace{0.2cm} \hat P_{\beta}=\mu \hat P_{\beta_c}- {\eta\over2\mu} \hat{\Omega}_{c}~,
\ea
where from now on the index $c$ denotes commutative variables, i.e. variables for which $\left[\hat{\Omega}_c, \hat{\beta}_c\right]=\left[\hat P_{\Omega_c}, \hat P_{\beta_c}\right]=0$ and $\left[\hat{\Omega}_c, \hat P_{\Omega_c}\right]=\left[\hat{\beta}_c, \hat P_{\beta_c}\right]=i$. The transformation Eqs. (\ref{eq1.9}) can be inverted, provided
\be\label{eq1.91}
\xi \equiv \theta \eta <1~.
\ee
The dimensionless constants $\lambda$ and $\mu$ are related by \cite{Bastos2},
\be\label{eq1.10}
\left(\lambda\mu\right)^2-\lambda\mu+{\xi\over4}=0\Leftrightarrow\lambda\mu={1+\sqrt{1-\xi}\over2}~.
\ee
The transformation Eqs. (\ref{eq1.9}) provide a representation of the operators (\ref{eq1.8.1}) as self-adjoint operators acting on the Hilbert space $L^2(\bkR^2)$. In this representation the WDW Eq. (\ref{eq1.8}) is deformed into a modified second order partial differential equation, which exhibits an explicit dependence on the noncommutative parameters
\be\label{eq1.11}
\left[-\left(-i \mu {\partial \over \partial {\Omega_c}}+{\eta\over2\mu}\beta_{c}\right)^2+\left(-i \mu {\partial \over \partial {\beta_c}}-{\eta\over2\mu}\Omega_c\right)^2-48\exp{\left[-2\sqrt{3}\left(\lambda\Omega_c+i{\theta\over2\lambda} {\partial \over \partial {\beta_c}} \right)\right]}\right]\psi(\Omega_c,\beta_c)=0~.
\ee

From Eq. (\ref{eq1.7})  we define a new constant operator $\hat A=\frac{\hat C}{\sqrt{1-\xi}}$, from which follows that
\be\label{eq1.12}
\mu \hat P_{\beta_c}+{\eta\over2\mu} \hat{\Omega}_c= \hat A~.
\ee
This new operator commutes with the noncommutative Hamiltonian of Eq. (\ref{eq1.11}). Hence we may look for solutions of Eq. (\ref{eq1.11}) that are simultaneous eigenstates of $\hat A$. A generic eigenstate of $\hat A$ with eigenvalue $a \in \bkR$, satisfies
\be\label{eq1.12a}
\left(-i\mu{\partial\over\partial\beta_c}+{\eta\over2\mu}\Omega_c\right)\psi_a(\Omega_c,\beta_c)=a \psi_a(\Omega_c,\beta_c)~.
\ee
which yields \cite{Bastos2}
\be\label{eq1.12b}
\psi_a(\Omega_c,\beta_c)=\Re_a(\Omega_c)\exp{\left[{i\over\mu}\left(a-{\eta\over2\mu}\Omega_c\right)\beta_c\right]}~.
\ee
Substituting this solution into Eq. (\ref{eq1.11}) implies that $\phi_a(z)\equiv\Re_a(\Omega_c(z))$ satisfies
\be\label{eq1.13}
\phi_a''(z)+\left(\eta z-a\right)^2\phi_a(z)-48\exp{[-2\sqrt{3}z+{\sqrt{3}\theta\over\lambda\mu}a]}\phi_a(z)=0~,
\ee
where we have introduced the new variable,
\be\label{eq1.14}
z={\Omega_c\over\mu}\hspace{0.2 cm}\rightarrow\hspace{0.2 cm}{d\over dz}=\mu{d\over d\Omega_c}~.
\ee
This is a second order ordinary differential equation that can be solved numerically. The equation itself depends on the eigenvalue $a$ and on the noncommutative parameters $\theta$ and $\eta$.

As shall be seen, Eq. (\ref{eq1.13}), can be used to obtain the temperature and the entropy of the Schwarzschild BH.

\section{Thermodynamics of Phase-Space Noncommutative Black Hole}

\begin{figure}
\begin{center}
\subfigure[ ~$\eta=0$ and $c=0.01$]{\includegraphics[scale=0.7]{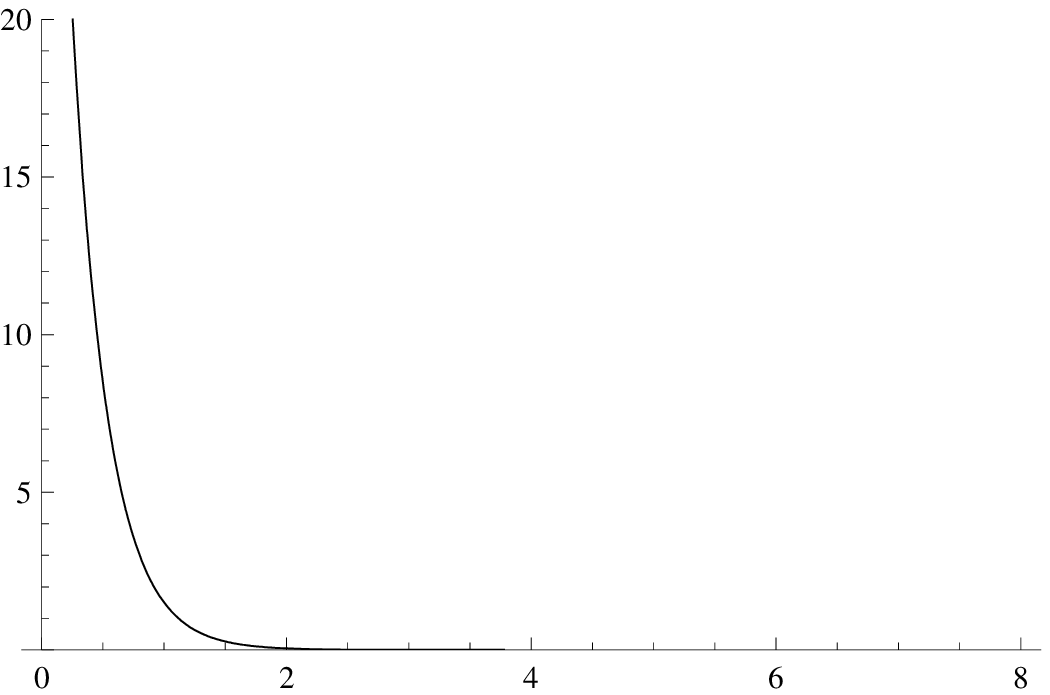}}
\subfigure[ ~$\eta=1.5$ and $c=5.68$]{\includegraphics[scale=0.7]{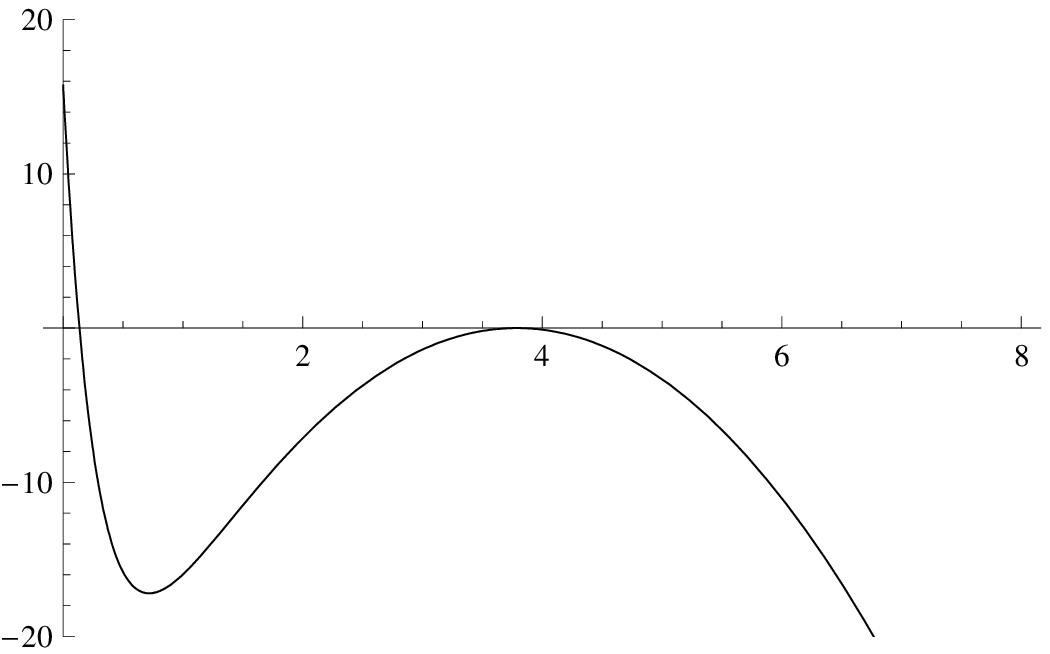}}
\caption{Potential function for some typical values of $\eta$ and $c$: (a) The potential function for the noncommutative case, $\theta\neq0$ and $\eta=0$ and (b) The potential function with $\eta\neq0$.  }
\label{fig:potential}
\end{center}
\end{figure}

We employ now the NCWDW equation to study the quantum behaviour of the interior of the Scharwzschild BH. Our aim is to obtain its temperature and entropy for the phase-space noncommutative extension. In order to get these quantities we compute the partition function through the Feynman-Hibbs procedure \cite{Obregon}. This method is based on the minisuperspace potential function, which can be determined from Eq. (\ref{eq1.13}):
\be\label{eq2.1}
V(z)=48\exp{[-2\sqrt{3}z+{\sqrt{3}\theta\over\lambda\mu}a]}-\left(\eta z-a\right)^2~.
\ee
For convenience, let us introduce a new variable,
\be\label{eq2.2}
x=z-{\theta\over2\lambda\mu}a~.
\ee
Thus, the potential function becomes,
\be\label{eq2.3}
V(x)=48\exp{(-2\sqrt{3}x)}-\left(\eta x-c\right)^2~,
\ee
where $c=P_{\beta}(0)+\eta\Omega(0)$ is a constant from the classical constraint. This potential function, which is depicted in Fig. \ref{fig:potential} b, exhibits a local minimum, $x_0$, and a local maximum.

In Fig. \ref{fig:potential} we present the potential function for the noncommutative case, $\theta\neq0$ and $\eta=0$, and the noncommutative case, where $\eta=1.5$. The case with noncommutativity only in the configuration variables (i.e. $\eta =0$, $\theta \ne 0$) \cite{Dominguez} is thus qualitatively similar to the commutative one, i.e. it has no local minimum. As can be seen in Eq. (\ref{eq2.1}), the noncommutative parameter $\theta$ appears in the argument of the exponential. Thus, unless noncommutativity in the momentum sector is present one does not have a local minimum of the potential for finite $x$. The values of $\eta$ used to plot the potential are fairly typical and the qualitative behaviour of the potential function is identical for other values. The constant $c$ is obtained using $P_{\beta}(0)=0.01$. The wave function, solution of the reduced NCWDW Eq. (\ref{eq1.13}), is well-defined for the chosen values \cite{Bastos2}.

Notice that the minimum of this potential is semiclassically stable. Following Ref. \cite{Coleman}, the probability of decay 
is given by $e^{-S_E}$ where $S_E$ is the euclidianized action for the bounce, i.e. for the motion of a 
particle travelling from $x_0=x(\tau_0)$ (the position where the potential attains its local minimum) to $x_1=x(\tau_1)$ (the position 
where it attains $V(x_1)=0$ after its local maximum) and arriving there with zero
velocity. After lifting the potential so that $V(x_0)=0$ (which does not affect the probability of decay) one finds that 
$S_E= \int_{x_0}^{x_1} \sqrt{2 V(x)}~dx$ \cite{Coleman}. Since during the 
relevant part of the motion the potential is 
essentially given by the second term in Eq. (\ref{eq2.3}), one can use values of Fig. \ref{fig:potential}, which are fairly typical, 
to obtain $S_E\simeq 9.5$. 

In order to compute the partition function using the Feynman-Hibbs procedure one resorts to the eigenvalue problem of the harmonic oscillator.
Thus we expand the exponential term in the potential Eq. (\ref{eq2.3}) to second order in powers of the  $x-x_0$ variable. The minimum $x_0$ is obtained by solving the following equation
\be\label{eq2.4}
{dV\over dx}|_{x_0}=-96\sqrt{3}\exp{(-2\sqrt{3}x_0)}-2 \eta (\eta x_0 - c)=0~.
\ee
The minimum is thus defined by the relation
\be\label{eq2.5}
\exp{(-2\sqrt{3}x_0)}=\zeta D-{\zeta^2\over\sqrt{3}}x_0~,
\ee
where $\zeta=\eta/4\sqrt{3}$ and $D=c/12$. Moreover, one should impose that the second derivative of the potential function is positive in order to obtain the following condition to the minimum,
\be\label{eq2.5a}
6\exp{(-2\sqrt{3}x_0)}-\zeta^2>0\Leftrightarrow x_0< -{1 \over \sqrt{3}} \ln \left({\zeta \over \sqrt{6}}\right) ~.
\ee
We then get for the potential function $V(x)$ to second order in $x-x_0$:
\be\label{eq2.6}
V(x)=48(6e^{-2\sqrt{3}x_0}-\zeta^2)(x-x_0)^2+48 e^{-2\sqrt{3}x_0} - (\eta x_0-c)^2~.
\ee
So, one can rewrite the ordinary differential equation resulting from the NCWDW equation as
\be\label{eq2.7}
-{1\over2}{d^2\phi\over dx^2}+24(6e^{-2\sqrt{3}x_0}-\zeta^2)(x-x_0)^2\phi+[24 e^{-2\sqrt{3}x_0} - {1 \over 2}(\eta x_0-c)^2]\phi=0~.
\ee

Comparing Eq. (\ref{eq2.7}) with the Schr\"odinger equation of the harmonic oscillator we can clearly identify the potential and the energy of the system. Thus, the noncommutative potential required to compute the partition function is
\be\label{eq2.8}
V_{NC}(y)=24(6e^{-2\sqrt{3}x_0}-\zeta^2)y^2~,
\ee
where we have defined a new variable $y=x-x_0$. The Feynman-Hibbs procedure allows for introducing quantum corrections to the partition function through the potential, i.e. one can admit a quantum correction for the potential given by \cite{Feynman},
\be\label{eq2.9}
{\beta_{BH}\over24}V_{NC}''(y)=2\beta_{BH}(6e^{-2\sqrt{3}x_0}-\zeta^2)~,
\ee
where $\beta_{BH}$ is the inverse of the BH temperature. The noncommutative potential with the corresponding quantum corrections then reads,
\be\label{eq2.10}
U_{NC}(y)=24(6e^{-2\sqrt{3}x_0}-\zeta^2)\left(y^2+{\beta_{BH}\over12}\right)~.
\ee

Finally, the noncommutative partition function is given by
\be\label{eq2.11}
Z_{NC}=\sqrt{1\over{48(6e^{-2\sqrt{3}x_0}-\zeta^2)}}{1\over\beta_{BH}}\exp{\left[-2{\beta_{BH}^2}\left(6e^{-2\sqrt{3}x_0}-\zeta^2\right)\right]}~.
\ee
We are now in conditions to compute the thermodynamic quantities of the Schwarzschild BH. Starting from the noncommutative internal energy of the BH, $\bar{E}_{NC}=-{\partial\over\partial\beta_{BH}}\ln Z_{NC}$,
\be\label{eq2.12}
\bar{E}_{NC}={1\over\beta_{BH}}+4(6e^{-2\sqrt{3}x_0}-\zeta^2)\beta_{BH}~,
\ee
the equality $\bar{E}_{NC}=M$, allows for obtaining an expression for the BH temperature:
\be\label{eq2.13}
\beta_{BH}={M\over8(6e^{-2\sqrt{3}x_0}-\zeta^2)}\left[1 \pm \left(1 - {16\over M^2}(6e^{-2\sqrt{3}x_0}-\zeta^2)\right)^{1/2}\right]~.
\ee
Inverting this quantity, one obtains the BH temperature. Dropping the term proportional to $M^{-2}$, as presumably $M>>1$, and considering the positive root, we obtain:
\be\label{eq2.13a}
T_{BH}={4 \over M}(6e^{-2\sqrt{3}x_0}-{\zeta^2})~.
\ee
Notice that this quantity is positive (cf. Eq. (\ref{eq2.5a})). To compare this result with the Hawking temperature $T_{BH}= {1 \over 8 \pi M}$, we should be cautious as the limit $\eta \rightarrow 0$ is ill-defined. Indeed, our whole construction rests on the premise that we have a stable local minimum of the potential. However, this is only true provided $\eta >0$. We remark that our expression displays the same mass dependence as the Hawking temperature, and we write it as
\be\label{eq2.13.1a}
T_{BH}={b(\zeta)\over M}~,
\ee
where $b$ is a $\zeta$ dependent constant:
\be\label{eq2.13.1b}
b(\zeta) = 4 (6e^{-2\sqrt{3}x_0}-{\zeta^2})~.
\ee
Moreover, a simple numerical calculation reveals that we may recover the Hawking temperature even in the presence of the momentum noncommutativity for a specific value of $\eta$. Indeed, equating Eq. (\ref{eq2.13a}) with the Hawking temperature and using the stationarity condition (\ref{eq2.5}), we obtain for $c=12 D =5.68$
\be\label{eq2.13.1c}
x_0=1.8478 \hspace{1 cm} \eta=0.025~.
\ee
And thus, since $\eta$ cannot be exactly equal to zero, we can regard $\eta_0=0.025$ as a reference value which yields the Hawking temperature, and as $\eta$ increases we get a gradual noncommutative deformation of the Hawking temperature.

The entropy is calculated using the definition, $S_{NC}=\ln Z_{NC}+\beta_{BH}\bar{E}_{NC}$. Thus, the entropy for the phase-space noncommutative BH is given by:
\ba\label{eq2.14}
S_{BH}&=&\ln{1\over\sqrt{12 b(\zeta)}}+{M^2\over{2 b(\zeta)}}\left(1+\sqrt{1-{4b(\zeta)\over M^2}}\right)+\nonumber\\
&-&{M^2\over{8 b(\zeta)}}\left(1+\sqrt{1-{4 b(\zeta)\over M^2}}\right)^2+\nonumber\\
&-&\ln{M\over{2 b(\zeta)}}\left(1+\sqrt{1-{4 b(\zeta)\over M^2}}\right)~.
\ea
Neglecting as before terms proportional to $M^{-2}$, as $M>>1$, we finally obtain
\be\label{eq2.16}
S_{BH}\simeq{M^2\over2 b(\zeta)}+\ln{\sqrt{b(\zeta)}\over{M \sqrt{3}}}~.
\ee
For the reference value $\eta=\eta_0$, we have $b(\zeta_0) = {1\over 8 \pi}$. We then recover the Hawking entropy, but also some "stringy" corrections
\be
S_{BH}=4\pi M^2+\ln{\sqrt{2 \pi\over3}}-\ln({8\pi M})~.
\ee\label{eq2.16a}
In summary, as in the case of the temperature, we obtain that the noncommutative entropy of the BH is the Hawking entropy plus additional contributions and some noncommutative corrections.

As can be clearly seen, the relevant thermodynamic quantities are corrections of the Schwarzschild BH ones that depend explicitly on the momentum noncommutative parameter $\eta$ through $b(\zeta)$. If $\eta= \eta_0$, the Hawking temperature and entropy for the black hole are recovered. On the contrary, by inspection, if $\eta=0$, the potential Eq. (\ref{eq2.1}) reduces to a monotonous exponential term with no local minima. It is only if $\eta\neq0$ that the potential function acquires a quadratic term. This in turn ensures that the potential function has a local minimum and a local maximum. It is around the minimum of the potential that the calculations of quantum corrections can be carried out. Thus, the Feynman-Hibbs procedure is meaningful only in the presence of the momentum noncommutativity.

\section{Singularity}

In this section, we use Eq. (\ref{eq1.11}) to extract some information about the essential $r=t=0$ singularity. It should be noticed that in order to employ the KS metric to describe the interior of the Schwarzschild black hole the identification (\ref{eq0.3}) has to be imposed. Thus, we can see that for $t=0$, $\Omega\rightarrow +\infty$ and $\beta\rightarrow +\infty$. We are thus interested in studying the limit
\be\label{eq3.1.1}
\lim_{\Omega_c,\beta_c \to +\infty} \psi(\Omega_c,\beta_c)~,
\ee
where $\psi(\Omega_c,\beta_c)$ is a generic solution of Eq. (\ref{eq1.11}). We seek for a representation of solutions of Eq. (\ref{eq1.11}) in terms of the eigenstates of $\hat A$ Eq. (\ref{eq1.12})
\be\label{eq3.1.2}
\psi(\Omega_c,\beta_c)= \int da \, C(a) \psi_a(\Omega_c,\beta_c)
\ee
where $C(a)\in \bkC$ and $\psi_a(\Omega_c,\beta_c)$ is of the form
\be\label{eq3.1.3}
\psi_a(\Omega_c,\beta_c)=\phi_a \left({\Omega_c\over\mu}\right) \exp\left[{i\over\mu}\left(a-{\eta\over2\mu}\Omega_c \right) \beta_c \right]
\ee
and $\phi_a(z)$, $z=\Omega_c/\mu$ satisfies Eq. (\ref{eq1.13}). In the limit $\Omega \rightarrow \infty$, if we keep only the dominant terms, Eq.(\ref{eq1.13}) reads:
\be\label{eq3.2}
\phi_a''(z)+\left(\eta z-a\right)^2\phi_a(z)=0~.
\ee
This equation can be rewritten for $\eta\neq0$ as
\be\label{eq3.3}
\left\{-{\partial^2\over\partial \tilde z^2}-\eta^2 \tilde z^2 \right\} \tilde\phi_a(\tilde z)=0~.
\ee
where the change of variables has been performed $\tilde z=z-\frac{a}{\eta}$ and $\tilde\phi_a(x)=\phi_a(x+\frac{a}{\eta})$. Eq. (\ref{eq3.3}) is analogous to the eigenvalue equation for an inverted harmonic oscillator. This Hamiltonian is self-adjoint in $L^2(\bkR^2)$, its spectrum is continuous and its zero eigenfunction (the solution of Eq. (\ref{eq3.3})) displays the asymptotic form \cite{Gitman1} (for $\eta \not=0$),
\be\label{eq3.4}
\tilde\phi_a(\tilde z)\sim {1\over\tilde z^{1/2}}\exp \left[\pm i {\eta\over2}\tilde z^2 \right]
\ee
and so, for all $a$,
\be\label{eq3.5}
\lim_{z \to +\infty} \phi_a(z) = \lim_{z \to +\infty} \tilde\phi_a(z-{a\over\eta})=0 \quad \Longrightarrow \quad \lim_{\Omega_c,\beta_c \to +\infty} \psi_a(\Omega_c,\beta_c)=0~.
\ee
Therefore, it seems reasonable to expect that, for a suitable, although fairly general choice of the coefficients $C(a)$
\be\label{eq3.6}
\lim_{\Omega_c,\beta_c \to +\infty} \psi(\Omega_c,\beta_c)=0~,
\ee
which is a necessary condition to provide a quantum regularization of the classical singularity of the Schwarzschild BH. However, one should be cautious before running into the conclusion that the probability of finding the BH at the singularity is zero. First of all the calculation of probabilities for general covariant systems is a subtle issue. In our case, given that the wave function is oscillatory in $\beta_c$, it seems natural to fix a $\beta_c$-hypersurface, corresponding to the introduction of the measure $\delta(\beta-\beta_c) d \beta d\Omega_c$ in the probability distributions. The probability $P(r=0,t=0)$ of finding the BH at the singularity would then be given by the expression:
\be\label{eq3.7}
P(r=0,t=0)= \lim_{\Omega_c,\beta_c \to +\infty} \int_{\Omega_c}^{+\infty}|\psi(\Omega_c',\beta_c)|^2 d \Omega_c'\simeq \lim_{\Omega_c \to +\infty}  \int_{\Omega_c}^{+\infty}|\phi_a({\Omega'_c\over\mu})|^2d\Omega_c'
\ee
which, unfortunately, is divergent. This follows from the conclusion (which can be derived from the asymptotic expression) that the inverted harmonic oscillator displays non-normalizable eigenstates \cite{Gitman1}. Hence, the noncommutativity of the form (\ref{eq1.8.1}) cannot be regarded as the final answer for the singularity problem of the Schwarzschild BH.

However, it is quite interesting to realize that an Hamiltonian $H$ with the asymptotic form
\be\label{eq3.8}
H= -{\partial^2\over\partial x^2} +V(x) \quad , \quad V(x) \sim -\eta^2 x^{2+2\epsilon}
\ee
for some $\epsilon >0$, displays zero energy eigenstates of the kind \cite{Gitman}
\be\label{eq3.9}
\psi(x) \sim {1\over x^{(1+\epsilon)/2}}\exp \left[\pm i \frac{\eta}{2+\epsilon}\tilde x^{2+\epsilon} \right]~,
\ee
which are normalizable. One may then speculate that a suitable perturbation of the noncommutative structure (\ref{eq1.8.1}) may lead to a NCWDW equation associated to a potential of the form $V(x)\sim -\eta^2 x^{2(1+\epsilon)}$ for some arbitrarily small $\epsilon >0$. The solutions of this new NCWDW equation would then display zero probability (in the sense of Eq. (\ref{eq3.7})) at the singularity, thus solving this problem for the Schwarzschild BH \cite{Bastos3}.

\section{Conclusions}

In this work, the NCWDW equation for a KS cosmological model was used to study the thermodynamic properties of the Schwarzschild BH and to obtain information about the structure of the singularity at $t=r=0$. Working out the NCWDW equation we implemented the Feynman-Hibbs procedure identifying the quadratic part of the minisuperspace potential. This allows for obtaining the partition function and hence the BH thermodynamic quantities. As we have shown, this is possible for a noncommutative BH only if noncommutativity in the momentum space is introduced.

The essential singularity $r=t=0$ is studied with the help of the NCWDW equation. In the limit of $t=0$, or $\Omega,\beta\rightarrow +\infty$, one arrives at the Schr\"odinger problem of the inverted harmonic oscillator, whose wave function vanishes at the limit $\Omega,\beta\rightarrow +\infty$, but is not square integrable. Therefore, it is not possible conclude that the singularity problem is solved by the considered momentum noncommutativity. Nevertheless, we can foresee that a noncanonical form of phase-space noncommutativity may allow for a square integrable wave function. This possibility will be analysed elsewhere \cite{Bastos3}.

\subsection*{Acknowledgments}

\vspace{0.3cm}

\noindent The authors would like to thank Guillermo Mena Marugan for the crucial discussion which has led our interest in the direction of black holes. The work of CB is supported by Funda\c{c}\~{a}o para a Ci\^{e}ncia e a Tecnologia (FCT) under the grant SFRH/BD/24058/2005. The work of NCD and JNP was partially supported by Grant No. PTDC/MAT/69635/2006 of the FCT.


\end{document}